\documentclass[12pt,draftclsnofoot,onecolumn]{IEEEtran}

\usepackage{cite}
\usepackage{graphicx}
\usepackage{balance}
\usepackage{color}
\usepackage{url}
\usepackage{graphicx, pstool}
\usepackage{float}
\usepackage{array}
\usepackage[english]{babel} 
\usepackage{amsmath,amssymb,amsfonts}
\usepackage{textcomp}
\usepackage{algorithmic}
\usepackage{caption}
\usepackage{breqn}
\usepackage{blindtext}
\usepackage{dblfloatfix} 
\usepackage[ruled,vlined,linesnumbered]{algorithm2e}
\makeatletter
\renewcommand{\fnum@figure}{Fig. \thefigure}
\makeatother

\newcolumntype{P}[1]{>{\centering\arraybackslash}p{#1}}
\newtheorem{theorem}{Theorem}

\newtheorem{lemma}{Lemma}

\newtheorem{corollary}{Corollary}

\newtheorem{proposition}{Proposition}

\newtheorem{conjecture}{Conjecture}

\newtheorem{sketch}{Sketch of Proof}

\begin{document}
	
	\title{Optimal Routing for Multi-user Multi-hop Relay Networks via Dynamic Programming}
	\author{
		Shalanika Dayarathna,~\IEEEmembership{Member,~IEEE,} Rajitha Senanayake,~\IEEEmembership{Member,~IEEE,} \\ and Jamie Evans,~\IEEEmembership{Senior Member,~IEEE.} 
	}	
	\maketitle
	
	\begin{abstract}
		In this paper, we study the relay selection problem in multi-user, multi-hop relay networks with the objective of minimizing the maximum outage probability across all users. When only one user is present, it is well known that the optimal relay selection problem can be solved efficiently via dynamic programming.  This solution breaks down in the multi-user scenario due to dependence between users. We resolve this challenge using a novel relay aggregation approach. On the expanded trellis, dynamic programming can be used to solve the optimal relay selection problem with computational complexity linear in the number of hops. Numerical examples illustrate the efficient use of this algorithm for relay networks.
	\end{abstract}
	
	
	\section{Introduction}\label{Sec-Intro}
	Cooperative transmission has been extensively studied as a promising paradigm for next generation wireless networks. In cooperative networks, relay nodes play an important role by acting as intermediate nodes that help combat fading, path loss and interference impairments \cite{electronics9030443}. For such networks, relay selection is critical in gaining advantages in terms of performance, complexity and overhead. As a result, much research has focused on relay selection (RS) in cooperative relay networks \cite{2809748,2970744,s18103263,3020299}. 
	
	In the literature, the multi-hop RS problem has been mainly analyzed for relay networks with a single user \cite{5982498,6364160,130145}. In \cite{5982498}, the authors show that an efficient dynamic programming approach implemented at a central processor can find the optimal relay path by comparing the bottleneck-link signal-to-noise-ratios (SNRs) of each path. When compared to exhaustive search, for a single-user decode-and-forward (DF) relay network, the optimal RS can be found very efficiently based on this approach. While, dynamic programming cannot be directly used in amplify-and-forward (AF) relay networks, it can be extended to a single-user AF relay network by approximating the effective SNR of a given path \cite{6364160,130145}. In \cite{6364160}, the authors propose a near optimal routing scheme based on the Viterbi algorithm by approximating the effective SNR of a given path as the minimum of SNRs of all the hops, an approximation which is tight at high SNR. Taking a different approach, in \cite{130145}, the authors express the outage probability in a recursive manner and propose a dynamic programming based algorithm for path selection. More recent work in this area focuses on machine learning techniques to select relay nodes in single-user, multi-hop relay networks \cite{electronics8090949,e23101310,2941932}. While these approaches provide interesting solutions in single-user networks, work on multi-user relay networks is still limited.
	
	With multiple source-destination (S-D) pairs, the RS problem becomes much more complicated, especially in DF relay networks. This is due to the potential interference and the limitation of one relay node serving one S-D pair to minimize the synchronization requirements. Even with no interference, the path selected by each user depends on the selection of other users. Therefore, a direct extension of the simple Viterbi algorithm is not possible for a multi-user relay network. Existing approaches to finding the optimal solution require searching through all possible paths from source nodes to destination nodes. This involves a computational complexity which is exponential in the number of hops and the number of S-D pairs. As a result, different sub-optimal RS strategies have been considered to offer less computational complexity \cite{130815}, decentralization \cite{2809748,3020299} and practical constraints \cite{s18103263}.
	
	In \cite{2091148}, the authors consider a dual-hop relay network and develop an optimal RS strategy to maximize the minimum SNR of all the users by using a linear marking mechanism. The RS scheme proposed in \cite{2091148} has a worst case complexity which is quadratic in the number of users and the number of relays. Taking a different approach, in \cite{130815}, the authors propose a sub-optimal RS scheme with linear complexity with respect to the number of relays for a multi-user dual-hop relay network. Later, this was extended to a multi-user, multi-hop relay network in \cite{2809748}, where the authors propose a decentralized RS scheme which performs RS in each hop independently, with the last two hops combined together to achieve full diversity. 
	
	With the dependence between S-D pairs, the existing dynamic programming based solutions for RS cannot be used in multi-user, multi-hop relay networks. To the best of our knowledge, there exists no low-complexity solution for optimal relay selection which provides a baseline for all these sub-optimal solutions. The naive or brute-force approach to the optimal relay selection involves exhaustively searching through all possible paths from source nodes to destination nodes. Therefore, in many of the previous works that provide sub-optimal solutions, the simulation results were either limited to smaller networks or performance was not compared against the optimal solution \cite{2809748}. We note that due to the connection between all the hops and users, the optimal relay selection in multi-user, multi-hop relay networks requires a centralized solution. The main contribution of this work is to provide a centralized solution that is guaranteed to obtain the optimal relay selection with linear complexity in terms of the number of hops. As such, our proposed solution provides a baseline that can be used to compare existing relay selection solutions to the optimal solution in larger relay networks, where the exhaustive search based optimal solution is not feasible.	
	
	\section{System Model}\label{Sec-model}
\begin{figure}
		\centering
		\includegraphics[width=0.75\textwidth]{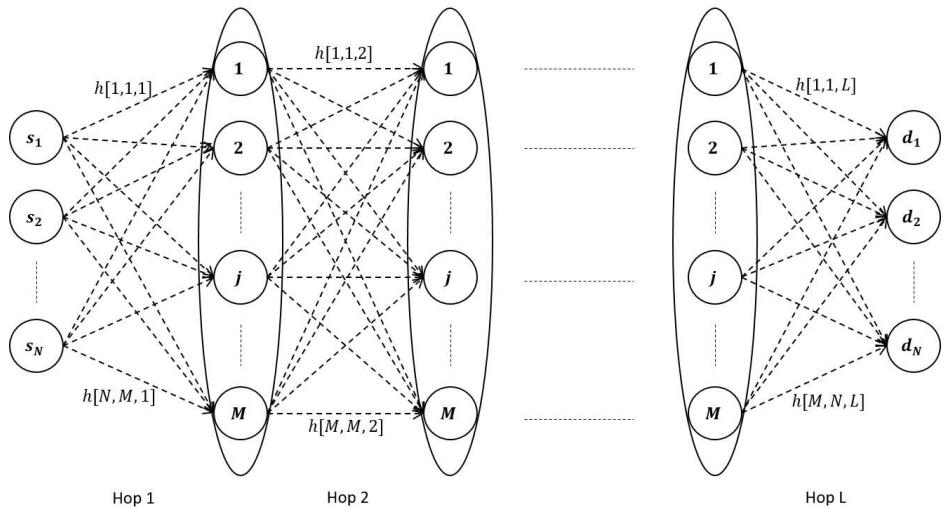}
		\centering\caption{A multi-user multi-hop relay network}
		\label{figure1}
	\end{figure}

	We consider a multi-user, multi-hop wireless relay network as illustrated in Fig. \ref{figure1}, where $N$ source nodes $(s_1,s_2,...,s_N)$ send information to $N$ corresponding destinations $(d_1, d_2,..., d_N)$. The communication is assisted by a multi-hop relay network consisting of $L$ hops with $M$ DF relays in each hop. Please note that even though we assume equal number of relay nodes in each hop for simplicity, this assumption is made without loss of generality. The more general case with different number of relay nodes in each hop can be handled by considering dummy relay nodes in each hop. For example, let us consider that the number of relay nodes are different in each hop and denote the number of relay nodes in hop $l$ by $M_l$. First, we define $M=\textrm{max} \{M_1,M_2,...,M_L\}$. Then we add $M-M_l$ dummy relay nodes in hop $l$ with zero channel gains for each link connected to these dummy nodes. As a result, there are $M$ relay nodes in each hop. In addition, due to zero SNR associated with dummy relay nodes, they will not be selected during the relay selection process.
	As commonly used in the literature, we assume that each S-D pair is assisted by only one relay in each hop and each relay assists at most one S-D pair to minimize the synchronization requirements, to avoid too much processing complexity in any single relay and to minimize power consumption in the network \cite{2809748,130815}. As one relay node supports at most one S-D pair, it is important to consider $M\ge N$ to ensure that all S-D pairs are able to communicate via the relay network. Therefore, we select $N$ relays from $M$ available relays in each hop and denote the relay selected for S-D pair $i$ in hop $l$ as $r(i,l)$.
	
	We model the channel gain between transmitter $i$ and receiver $j$ in hop $l$ as a random variable $h[i,j,l]$. In general, this includes small scale fading, path loss and shadowing. We also assume that each node operates in half-duplex mode with transmit power $P$ and that transmissions are scheduled so that cross-hop (or inter-hop) interference can be neglected \cite{2809748,6364160}. For such a network, the received signal at node $j$ in hop $l$ can be written as,
	\begin{align}
	y[j,l] = \sum_{i=1}^{N}\sqrt{P}h[r(i,l-1),j,l]\; x[r(i,l-1),l-1] + n[j,l]
	\end{align}
	where $x[r(i,l-1),l-1]$ is the information symbol transmitted by node $r(i,l-1)$ in hop $l-1$ which has unit average energy and $n[j,l]$ is the additive white Gaussian noise at node $j$ in hop $l$ with mean zero and variance $\sigma^2$. 
	
	\section{Optimal Relay Selection}\label{Sec-optimal}	
	The main challenge in a multi-user, multi-hop relay network is selecting the best relay combination for each S-D pair. As each relay node only supports one S-D pair, the RS of one S-D pair affects the RS of other S-D pairs. In addition, the signal-to-interference-plus-noise-ratio (SINR) of each link in hop $l, \forall l > 1$ depends on the transmitting relay nodes in that hop which in turn depend on the RS of the previous hop. As such, the optimal relay selection in any given hop depends on the relay selection of every other hop.
	In a relay network with $L$ hops where each hop has $M$ relays, there are $M^{L-1}$ possible paths for each S-D pair. Therefore, a path can be thought of as the set of relays that assists the communication between the source node and the relevant destination node. In a multi-user relay network with $N$ S-D pairs, there are $\prod_{i=0}^{N-1} (M-i)^{L-1}$ possible non-overlapping paths. The optimum RS can be solved by selecting $N$ paths from these non-overlapping paths such that the performance objective is optimized. 
	
	In this paper, we consider the RS problem for a multi-user, multi-hop relay network with multiple DF relays by focusing on the network outage probability performance. For a given path the end-to-end received SINR of S-D pair $i$ depends on the minimum SINR over all the hops and can be expressed as,
	\begin{align}\label{eq_1}
	\gamma_i = \textrm{min}\{\gamma_{i,1},\gamma_{i,2},...,\gamma_{i,l},...,\gamma_{i,L}\},
	\end{align}	
	where $\gamma_{i,l}$ is the received SINR corresponding to S-D pair $i$ in hop $l$ and can be expressed as,
	\begin{align}\label{eq_2}
		\gamma_{i,l} = 
		\dfrac{P|h[r(i,l-1),r(i,l),l]|^2}{\sigma^2+\sum_{j \neq i}^N P|h[r(j,l-1),r(i,l),l]|^2},
	\end{align} 
	with $r(i,0)=s_i$ and $r(i,L)=d_i$. We note that an outage occurs for S-D pair $i$ when its end-to-end received SINR is less than the required threshold $\gamma_{th}(i)$. Therefore, the outage probability of a given S-D pair $i$ can be calculated as \cite{5982498},
	\begin{align}\label{eq3}
	P_{out}(i) = \textrm{Pr}\bigg[\gamma_i < \gamma_{th}(i)\bigg].
	\end{align}
	Let us define $\bar{\gamma_i}=\dfrac{\gamma_i}{\gamma_{th}(i)}$ to denote the normalized end-to-end received SINR of S-D pair $i$ with respect to its required threshold. As such, \eqref{eq3} can be simplified into
	\begin{align}\label{eq4}
	P_{out}(i) = \textrm{Pr}\bigg[\bar{\gamma_i} < 1\bigg].
	\end{align}
	In this work, we consider that an outage occurs in the network when any S-D pair is in outage. As such, we define the network outage as,
	\begin{align}\label{eq5}
	P_{out} &= 1-\textrm{Pr}\bigg[\bar{\gamma_1} \ge 1,\bar{\gamma_2} \ge 1,...,\bar{\gamma_i} \ge 1,...,\bar{\gamma_N} \ge 1\bigg] \nonumber \\
	&= 1-\textrm{Pr}\bigg[{\underset{i \in \{1,...,N\}} {\textrm{min}}}\{\bar{\gamma_i}\} \ge 1\bigg].
	\end{align}
	Therefore, we can minimize $P_{out}$ by maximizing $\textrm{Pr}\bigg[{\underset{i \in \{1,...,N\}} {\textrm{min}}}\{\bar{\gamma_i}\}\ge 1\bigg]$ which in turn is equivalent to
	\begin{align}\label{eq6}
	{\underset{r(i,l), \forall i, l} {\textrm{max}}}\bigg({\underset{i \in \{1,...,N\}} {\textrm{min}}}\{\bar{\gamma_i}\}\bigg),
	\end{align}
	where $\bar{\gamma_i}$ depends on $r(i,l)$ as given in \eqref{eq_2}. Therefore, in this work, we focus on finding the relay path that maximizes the minimum normalized end-to-end received SINR among all S-D pairs with respect to their required threshold values, which in fact minimizes the network outage. In the rest of the paper, we refer $\bar{\gamma_i}$ as the end-to-end received SINR of S-D pair $i$ for simplicity.
	
	We note that when the number of users is limited to one, the optimal path can be found using dynamic programming techniques such as the Viterbi based algorithm where a central controller can be used to select the path that has the largest bottleneck-link SNR \cite{5982498}. However, in the presence of multiple users, this solution breaks down due to the dependence between users. As such, the optimal solution is obtained by exhaustively searching through all possible paths from source nodes to destination nodes. This approach has an exponential complexity with respect to $N$ and $L$ as there are $\prod_{i=0}^{N-1} (M-i)^{L-1}$ possible paths for $N$ S-D pairs \cite{2809748}. Therefore, calculating all possible paths from source nodes to destination nodes can be a very complex task in large networks. 
	
	In the following, we reformulate the optimal relay selection problem by mapping the multi-user relay network into a trellis diagram in which the goal is to find the path that maximizes the minimum branch weight. We then propose an efficient centralized path selection algorithm for multi-user DF relay networks that can achieve the optimal relay assignment with linear complexity with respect to the number of hops in the network.
	
	\section{Trellis based Optimal Relay Selection}
\begin{figure}
		\centering
		\includegraphics[width=0.8\textwidth]{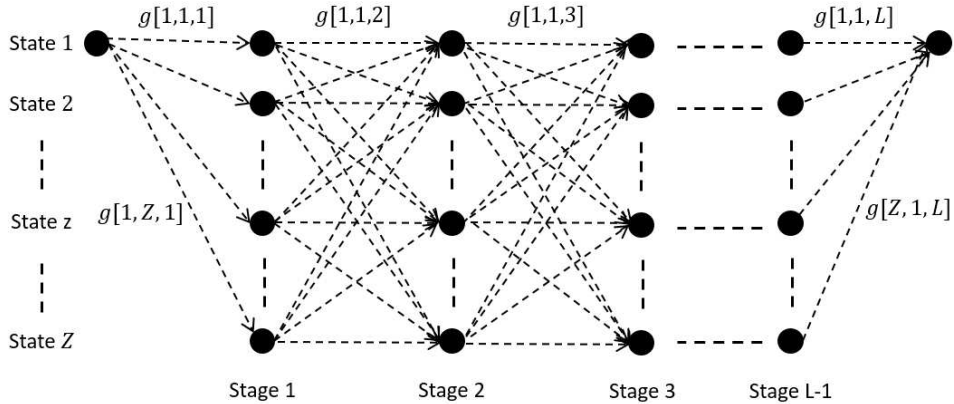}
		\centering\caption{Mapped trellis for the multi-hop relay network}
		\label{figure2}
	\end{figure}

	A trellis diagram consists of a finite number of stages between the starting point and the end point. Each stage consists of a finite set of states. In addition, each state in a given stage is connected to every other state in the next stage via a branch. Noting that there are $Z=\prod_{i=0}^{N-1}(M-i)$ possible relay combinations in each hop, we consider a trellis diagram of $L$ stages with $Z$ states in each stage, as illustrated in Fig.~ \ref{figure2}. As such, the number of stages in the trellis diagram can be mapped to the number of hops in the network and each state in the trellis will be similar to a possible relay combination in each hop. Noting that each state consists of a sequence of ordered $N$ nodes, the branch between state $i$ in stage $l-1$ and state $j$ in stage $l$ includes $N$ links. These links represent the communicating links of $N$ S-D pairs when the ordered node sequence given by state $i$ transmits to the respective ordered node sequence given by state $j$ in hop $l$. Therefore, we define the branch weight between state $i$ in stage $l-1$ and state $j$ in stage $l$ as the minimum received SINR among these $N$ links and denote this by $g[i,j,l]$. With this model, we map the trellis start and end points to source and destination combinations given by $\{s_1,s_2,...,s_N\}$ and $\{d_1,d_2,...,d_N\}$, respectively. Therefore, each path in the trellis represents one of the possible relay paths in the network.
	
	As the optimal relay selection maximizes the minimum end-to-end received SINR for all the S-D pairs, the effective SINR for a given path is the minimum SINR over all the S-D pairs and hops in that path. Noting that the branch weight is mapped to the minimum SINR among all the S-D pairs in a given hop, finding the trellis path that maximizes the minimum branch weight provides the optimal relay path. For a trellis with $L$ stages and $Z$ states in each stage, there are $Q=Z^{L-1}$ paths. Let us define $f(l,q)$ as the branch weight between stage $l-1$ and stage $l$ for a given trellis path $q$ and formulate the optimization problem as,
	\begin{align}\label{eq_6}
	&{\underset{q\in\{1,...,Q\}}{\textrm{max}}}\;\bigg({\underset{l \in \{1,...,L\}} {\textrm{min}}}\{f(l,q)\}\bigg). 
	\end{align}
	Based on the trellis, we next propose a dynamic programming based efficient path selection algorithm. The main focus on this paper is to find a low-complexity optimal relay selection solution and as commonly assumed in literature, we assume that full channel state information (CSI) is available at a central control node and it is in charge of RS decision \cite{electronics9030443, 2809748, 2941932}. At this control node, branch weights $g[i,j,l]$ are computed based on global CSI and given as input to the proposed algorithm along with all possible relay combinations $Z$ and the number of hops $L$.
	
	\subsection{Dynamic Programming based Path Selection Algorithm}\label{Sec-algo} 
	Dynamic programming is performed by breaking down the complex problem into simpler sub-problems and then finding the optimal solution to these sub-problems \cite{130145}. Please note that, if the optimal path contains state $z$ in stage $l$, then there cannot be a better path that leads up to state $z$ even if we break the trellis diagram at stage $l$. As such, the optimal path in the trellis can be obtained by considering the optimal path up to each state in a given stage. Therefore, in the proposed algorithm, we break the optimization problem given in \eqref{eq_6} to several sub-problems where each sub-problem seeks to maximize the minimum branch weight up to a given state in a given stage. 
	
	Let us denote the maximum branch weight up to state $z$ in stage $l$ as $u(z,l)$. Next, we can show that $u(z,l)$ can be computed recursively as,
	\begin{align}\label{eq_5}
		u(z,l)={\underset{\bar{z} \in \{1,...,Z\}} {\textrm{max}}}\biggl(\textrm{min}\biggl\{g[\bar{z},z,l], u(\bar{z},l-1)\biggr\}\biggr),
	\end{align}
	with $u(z,1)=g[1,z,1]$. Please note that we can obtain the optimal path up to state $z$ in stage $l$ by finding the trellis path that maximizes $u(z,l)$. As such, the sub-problem at state $z$ in stage $l$ is given by \eqref{eq_5}. We note that while the optimization given in \eqref{eq_5} results in the optimal relay selection, it is not directly connected to the optimization problem of maximizing the minimum end-to-end received SINR over all S-D pairs.
	\begin{algorithm}
		\DontPrintSemicolon 
		\SetAlgoLined
		\SetKwInOut{Input}{Input}\SetKwInOut{Output}{Output}
		\Input{Branch weights, $Z, L$}
		\Output{Optimum state vector $\mathbf{x_{opt}}$}
		
		$\mathbf{U}=0,\; \mathbf{V}=0,\; z^*=1$\;
		$u(z,1) = g[1,z,1],\, v(z,1) = 1, \, \forall z\in\{1,...,Z\}$ \;
		$u(z,l) \!=\! {\underset{\bar{z} \in \{1,...,Z\}} {\textrm{max}}}\biggl(\!\textrm{min}\biggl\{g[\bar{z},z,l], u(\bar{z},l-1)\biggr\}\biggr), $\;
		$v(z,l) \!=\! {\underset{\bar{z} \in \{1,...,Z\}} {\textrm{argmax}}}\biggl(\!\textrm{min}\biggl\{g[\bar{z},z,l], u(\bar{z},l-1)\biggr\}\biggr),$ 
		\nonumber{$ \forall z\in\{1,...,Z\}, l\in\{2,...,L-1\}$}\;
		$u(1,L) = {\underset{\bar{z} \in \{1,...,Z\}} {\textrm{max}}}\biggl(\textrm{min}\biggl\{g[\bar{z},1,L], u(\bar{z},L-1)\biggr\}\biggr)$\;
		$v(1,L) = {\underset{\bar{z} \in \{1,...,Z\}} {\textrm{argmax}}}\biggl(\textrm{min}\biggl\{g[\bar{z},1,L], u(\bar{z},L-1)\biggr\}\biggr)$\;
		\For{$l=L:-1:2$}{
			$z^*=v(z^*,l)$\;
			$\mathbf{x_{opt}} (l-1) = z^* $ \;
		}
		\caption{The Proposed Optimal RS Algorithm}
		\label{Algorithm_1}
	\end{algorithm}
	
	In step 1 of Algorithm \ref{Algorithm_1}, we define and initialize the local variable $z^*$ as one. We define two zero matrices $\mathbf{U}$ and $\mathbf{V}$, and denote the $(z,l)^{\textrm{th}}$ element of $\mathbf{U}$ and $\mathbf{V}$ by $u(z,l)$ and $v(z,l)$, respectively, where $z$ refers to the state and $l$ refers to the stage.
	Then in step 2, we consider the first stage and for each state $z$ we assign the branch weight $g[1,z,1]$ to $u(z,1)$ and number $1$ to $v(z,1)$. Next, we consider the second stage and for each state $z$, we first consider each state $\bar{z}$ in the first stage and calculate the minimum of $u(\bar{z},1)$ and $g[\bar{z},z,2]$. Then we maximize over all the states in the first stage as in \eqref{eq_5} and assign that maximum value to $u(z,2)$ and the relevant index to $v(z,2)$. We continue this for all the states in all the stages in step 3-6. Then, we start with the last stage and update $z^*$ with $v(1,L)$. In step 9, we assign $z^*$ to $(L-1)^{\textrm{th}}$ element of the optimum state vector $\mathbf{x_{opt}}$ denoted by $\mathbf{x_{opt}}(L-1)$. Then in step 8, we consider the previous stage and update $z^*$ with $v(z^*,L-1)$. Likewise, we continue tracking the best state for each stage until we reach the first stage. At the end of Algorithm \ref{Algorithm_1}, $l^{\textrm{th}}$ element of $\mathbf{x_{opt}}$ represents the optimal state in stage $l$ for the optimal trellis path. In other words, $\mathbf{x_{opt}}(l)$ contains the index of the selected relay combination in hop $l$ of the optimal relay assignment. 
	
	\subsection{Complexity Analysis}\label{Sec-complex} 
	In this work, we map the multi-user, multi-hop relay network into a trellis diagram with $L$ stages and $Z$ states in each stage. Solving for the best path based on Algorithm \ref{Algorithm_1}, which uses dynamic programming, involves $Z+Z^2(L-2)$ comparisons and $2+2Z(L-1)$ memory elements. Therefore, in optimal relay selection, the proposed algorithm has a linear complexity with respect to the number of hops $L$, which is significantly better than the alternative exhaustive search approach which has an exponential complexity.	
	
	\section{Simulation Results}\label{Sec-sim}
	In this section, we provide some numerical examples to illustrate the use of the proposed optimal relay selection algorithm. 
	
	Unless otherwise specified, we consider a relay network where the channels between nodes are subject to large-scale fading, which is modeled in-terms of distance dependent path loss with a path loss exponent $3.6$ which corresponds to urban areas \cite{pathLossExponent} and log-normal shadowing with standard deviation $8$, and small-scale fading which is modeled in-terms of Rayleigh fading with zero mean and unit variance. We compute the path loss between nodes by assuming equal distance between hops with the distance between source nodes and destination nodes set to $D_L$. We assume that all nodes have the same transmit power $P$ which varies between $16$ dBm--$40$ dBm ($\approx 16$ mW--$10$ W). In many of the previous works on multi-user, multi-hop relay networks, the simulation results were limited to smaller networks \cite{2809748}. This was due to the complexity involved in the optimal RS via exhaustive search. In the following we illustrate that our approach does not have such a limitation, hence allowing us to illustrate performance for larger relay networks. The simulation results are generated using Monte-Carlo simulations where the small-scale fading is averaged over $10^5$ time slots. On the other hand, we consider that the large-scale fading is fixed for the duration of the simulation.
	
\begin{figure}
		\centering
		\includegraphics[width=0.75\textwidth]{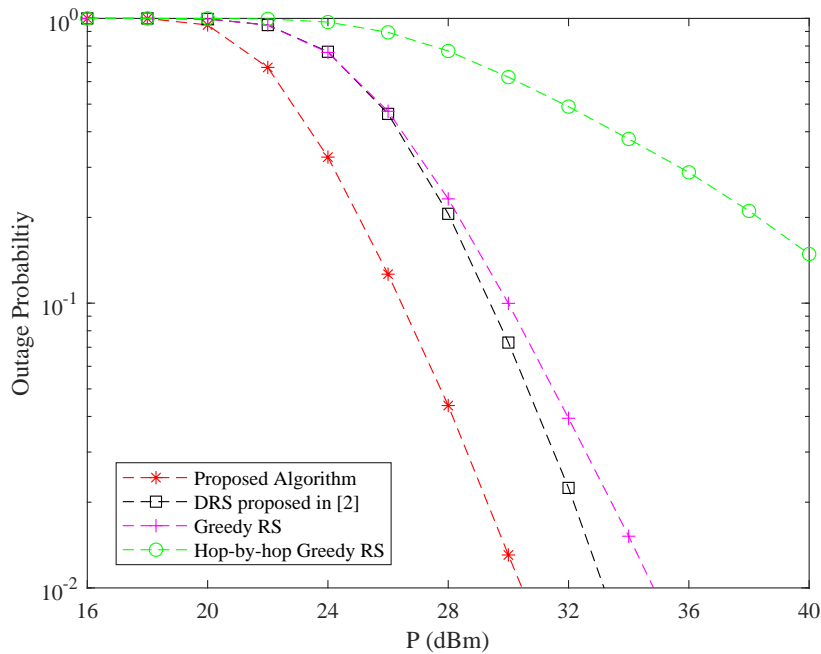}
		\centering\caption{Outage probability for multi-user networks with $N=3, M=4, L=6$ and $D_L=3$ km}
		\label{figure6}
	\end{figure}

	Fig. \ref{figure6} plots the network outage probability versus the transmit power for the proposed RS in comparison with three existing RS schemes, namely, the decentralized RS (DRS) proposed in \cite{2809748}, the greedy RS solution and hop-by-hop greedy RS solution. We consider a multi-user network $(N = 3)$ with $M = 4, L = 6$ and $D_L = 3$ km. Under the greedy RS, the first user selects the best relay path over all the hops. Then the second user selects the best possible relay path over all the hops from the remaining set of relays and so on. As such, the greedy approach does not maintain user fairness. Under the hop-by-hop greedy RS, we consider greedy RS implemented in hop-by-hop manner. We note that the greedy RS performs best when there is no interference between S-D pairs. As such, to perform a fair comparison, we consider that there is no interference between S-D pairs and set the SNR threshold to $3$ dB. From the plot, we observe that the solutions based on greedy RS result in worse outage probability compared to the optimal RS and the DRS proposed in \cite{2809748}. We can also observe that even for smaller number of hops and S-D pairs, the DRS proposed in \cite{2809748} has a considerably higher network outage than the optimal RS due to its hop-by-hop approach which removes some of the possible relay combinations. As such, we can expect the gap between the proposed algorithm and existing RS solutions to increase further when either the number of hops or the number of S-D pairs increases.
	
	As discussed in section \ref{Sec-complex}, the complexity of the proposed algorithm is linear in the number of hops, while it remains exponential in the number of S-D pairs. The DRS proposed in \cite{2809748} has a complexity that is linear in the number of hops and quadratic in the number of S-D pairs. On the other hand, both greedy solutions have a complexity that is linear in both the number of hops and the number of S-D pairs. The average computational times and the outage probability for the four relay selection solutions are presented in Table \ref{table1} and Table \ref{table2}, respectively. We consider a multi-user, multi-hop relay network with $M=6$ and $D_L = 5$ km. From Table \ref{table1}, we can clearly see that the computational time of existing relay selection solutions and the proposed optimal relay selection increases linearly with $L$. On the other hand, computational time of the proposed optimal RS increases exponentially with the number of S-D pairs $N$.  For the decentralized relay selection proposed in \cite{2809748} the computational time increases quadratically and that of the two greedy solutions increases linearly with $N$. From Table \ref{table2}, we can also observe that the difference between the outage probability of the existing relay selection solutions and the optimal relay selection increases significantly with both $L$ and $N$, thus increasing their sub-optimality. Therefore, we can clearly see that the proposed optimal relay selection obtains better outage performance with increasing $L$ while maintaining a linear complexity. On the other hand, with increasing $N$, there is a clear trade-off between the complexity in terms of the number of S-D pairs and the outage performance.
	\begin{table}
		\centering
		\caption{Computational Time in milliseconds when $M=6$ and $D_L=5$ km}
		\begin{tabular}
			{ |p{3cm}|p{2.75cm}|p{2.75cm}|p{2.75cm}|p{2.75cm}| }
			\hline
			& Optimal RS & DRS proposed in \cite{2809748} & Greedy Solution & Hop-by-hop Greedy Solution \\ \hline  
			$N=2, L=8$ & 2.8 & 0.1000 & 0.5056 & 0.04343 \\ \hline		
			$N=2, L=10$ & 3.3 & 0.1186 & 0.5769 & 0.05319 \\ \hline
			$N=2, L=12$ & 3.8 & 0.1367 & 0.6452 & 0.06293 \\ \hline
			$N=2, L=14$ & 4.4 & 0.1564 & 0.7245 & 0.07283  \\ \hline
			$N=2, L=10$ & 3.3 & 0.1 & 0.5 & 0.05319 \\ \hline		
			$N=3, L=10$ & 28.4 & 0.2 & 0.8 & 0.06892 \\ \hline	
			$N=4, L=10$ & 232.1 & 0.7 & 0.11 & 0.08544 \\ \hline	
			$N=5, L=10$ & 944.0 & 5.3 & 0.14 & 0.09993 \\ \hline
		\end{tabular}
		\label{table1}
	\end{table}	
	\begin{table}
		\centering
		\caption{Outage Probability when $M=6$ and $D_L=5$ km}
		\begin{tabular}
			{ |p{3cm}|p{2.75cm}|p{2.75cm}|p{2.75cm}|p{2.75cm}| }
			\hline
			& Optimal RS & DRS proposed in \cite{2809748} & Greedy Solution & Hop-by-hop Greedy Solution \\ \hline  
			$N=2, L=8$ & 0.6074 & 0.9607 & 0.7073 & 0.9878 \\ \hline
			$N=2, L=10$ & 0.0206 & 0.2219 & 0.0354 & 0.6997 \\ \hline
			$N=2, L=12$ & 0.0007 & 0.0155 & 0.0018 & 0.4289 \\ \hline
			$N=2, L=14$ & 0.00001 & 0.0010 & 0.0002 & 0.2714 \\ \hline
			$N=2, L=10$ & 0.0206  & 0.2219 & 0.0354 & 0.6997 \\ \hline
			$N=3, L=10$ & 0.0315 & 0.3332 & 0.1057 & 0.8639 \\ \hline
			$N=4, L=10$ & 0.0443 & 0.4509 & 0.3501 & 0.9570 \\ \hline
			$N=5, L=10$ & 0.0782 & 0.5856 & 0.8958 & 0.9951 \\ \hline
		\end{tabular}
		\label{table2}
	\end{table}

\begin{figure}
		\centering\includegraphics[width=0.75\textwidth]{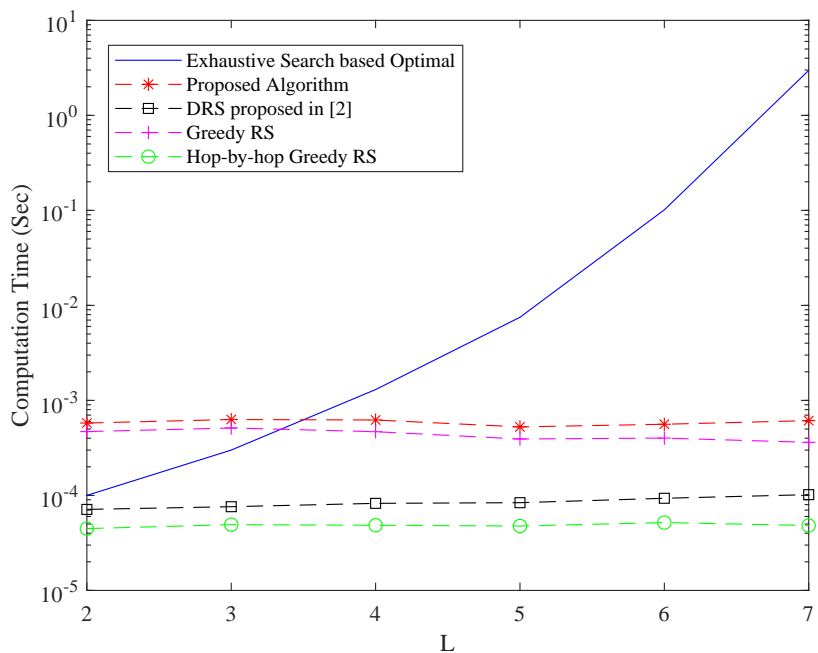}
		\centering\caption{Average computation time versus $L$ for a two-user network with $M=3$ and $D_L=3$ km}
		\label{figure7}
	\end{figure}

	Fig. \ref{figure7} plots the average computational time versus the number of hops $L$ for a two-user network ($N=2$) with $M=3$ and $D_L=3$ km. From the plot, we can observe that the average computational time of the exhaustive search based optimal solution increases linearly in the log scale indicating that it has an exponential complexity with respect to $L$. On the other hand, proposed algorithm and the existing sub-optimal relay selection solutions have average computational time which fluctuates around a constant in the log scale, thus indicating their linear complexity with respect to $L$.
	
\begin{figure}
		\centering
		\includegraphics[width=0.75\textwidth]{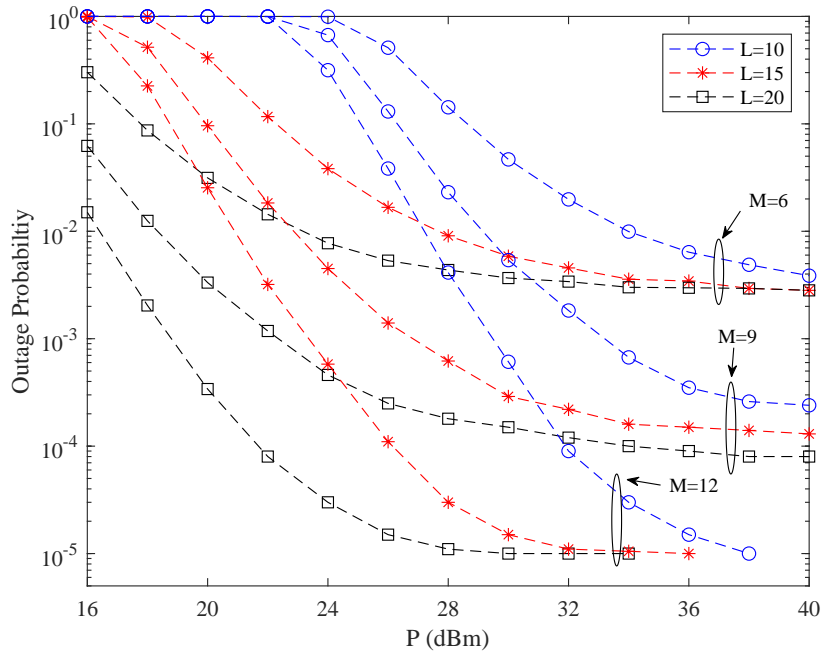}
		\centering\caption{Outage probability for multi-user networks with $N=2, M=6,9,12, L=10,15,20$ and $D_L=10$ km}
		\label{figure4}
	\end{figure}

	Next, we consider that there is interference between transmissions in the same hop. We also consider only the path loss for large-scale fading. Fig. \ref{figure4} plots the network outage probability versus the transmit power for two-user networks ($N=2$) with $M=6,9,12, L=10,15,20$ and $D_L=10$ km when SINR threshold $\gamma_{th}$ is set to $-3$ dB. From the plot, we observe that outage probability decreases with increasing $M$. As $M$ increases, the number of available relay combinations increases. As a result, the probability of each S-D pair selecting a relay with larger gain and lower interference increases, thus increasing the end-to-end received SINR of each S-D pair. Therefore, when we have more relays available in each hop, the overall outage probability reduces. A similar observation can be made with respect to $L$. As $L$ increases, the distance between transmitting and receiving nodes of a given hop decreases. This reduces the path loss between two nodes, thus increasing the received SINR. We also observe that in the high transmit power regime, the outage probability of a given $M$ enters a saturation with increasing $L$. This can be explained by the interference dominance in the high transmit power regime, because the reduction in path loss increases both the received SNR and the received interference.	
	
\begin{figure}
		\centering
		\includegraphics[width=0.75\textwidth]{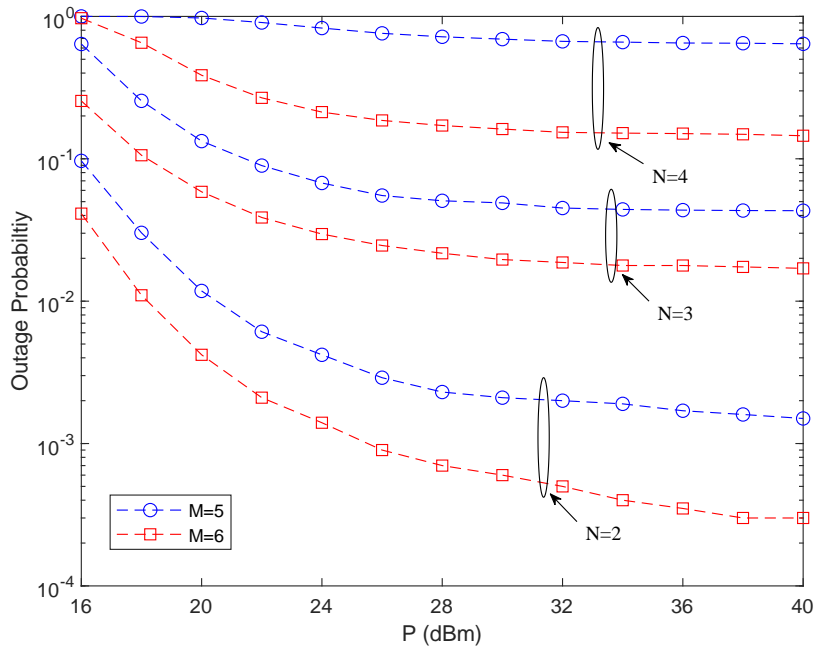}
		\centering\caption{Outage probability for multi-user networks with $N=2,3,4, M=5,6, L=20$ and $D_L=10$ km}
		\label{figure5}
	\end{figure}

	Fig. \ref{figure5} plots the network outage probability versus the transmit power for a multi-user network ($N=2,3,4$) with $M=5,6, L=20$ and $D_L=10$ km when the SINR threshold is set to $-5$ dB. From the plot, we observe that as the number of S-D pairs $N$ increases, the overall outage probability increases due to the effect of interference. With more S-D pairs, the number of available relay combinations decreases whereas the impact of interference increases in each hop. As a result, the end-to-end received SINR of each S-D pair decreases, thus increasing the outage probability. We also observe that unlike with $L$, the outage probability of a given $M$ does not enter a saturation with increasing $N$ in the high transmit power regime. With $M >> N$, a given S-D pair has more paths to traverse through the network. As such, there is a high probability for every S-D pair to select a relay path with high end-to-end received SINR under different channel fading, thus minimizing the outage probability.
	
	\section{Conclusion}\label{Sec-Plans}
	We considered a DF multi-hop relay network with multiple users and multiple relays in each hop. By introducing a set of states that aggregate the possible relay combinations among users, we showed that the network topology can be mapped to an expanded trellis diagram. Based on that, we proposed an optimal path selection algorithm that maximizes the minimum end-to-end received SINR of all users. Our proposed algorithm has a linear complexity with respect to the number of hops, thus, operates with much less processing power when compared to the exhaustive search that has an exponential complexity with respect to the number of hops. The proposed approach can be used to find the optimum RS in networks with larger number of relays and hops. While current work focus on the relay selection problem, the joint analysis of both power control and relay selection would be a desirable future extension.
	
	\section*{Acknowledgments}
	The authors would like to thank Margreta Kuijper and Peter Dower for useful discussions and insights which motivated this paper.

	
	\bibliography{References}
	\bibliographystyle{IEEEtran}
	
\end{document}